\documentclass[english,12pt,a4paper]{article}
\pdfoutput=1
\usepackage{authblk}
\usepackage[text={17.0cm,23.7cm}]{geometry}
\usepackage[T1]{fontenc}
\usepackage[utf8]{inputenc}
\usepackage{babel}
\usepackage{hyperref}
\usepackage{cite}
\usepackage{graphicx}
\usepackage{amsmath}
\usepackage{amssymb}
\usepackage[dvipsnames]{xcolor}
\usepackage{yfonts}
\usepackage{ulem}
\usepackage{enumitem}
\usepackage[mathscr]{euscript}
\usepackage{float}

\begin{document}
	\title{\vspace{-2cm}
		{\normalsize
			\flushright TUM-HEP 1320/21\\}
		\vspace{0.6cm}
		\textbf{Direct detection of non-galactic light dark matter}\\[8mm]}
	\author[1,2]{Gonzalo Herrera \footnote{\href{mailto:gonzalo.herrera@tum.de, gonzalo.herrera@tum.de}{gonzalo.herrera@tum.de}}}
	\author[1]{Alejandro Ibarra \footnote{\href{mailto:ibarra@tum.de}{ibarra@tum.de}}}
	\affil[1]{\normalsize\textit{Physik-Department, Technische Universit\"at M\"unchen}\\\textit{James-Franck-Stra\ss{}e, 85748 Garching, Germany}}
	\affil[2]{\normalsize\textit{Max-Planck-Institut für Physik (Werner-Heisenberg-Institut), Föhringer Ring 6,80805 M\"unchen, Germany}}
	
	\date{}
	
	\maketitle
	\begin{abstract}
	A fraction of the dark matter in the solar neighborhood might be composed of non-galactic particles with speeds larger than the escape velocity of the Milky Way. The non-galactic dark matter flux would enhance the sensitivity of direct detection experiments, due to the larger momentum transfer to the target.
	In this note, we calculate the impact of the dark matter flux from the Local Group and the Virgo Supercluster diffuse components in nuclear and electron recoil experiments. The enhancement in the signal rate can be very significant, especially for experiments searching for dark matter induced electron recoils. 
	\end{abstract}

\section{Introduction}

 Various astronomical and cosmological observations point towards the existence of dark matter in our Universe, possibly constituted by a population of new elementary particles. A consequence of this hypothesis is that the Earth should be constantly bombarded by dark matter particles. Although the particle nature of dark matter is still unknown, it is plausible that the dark matter particle could couple to the Standard Model sector through other interactions aside from gravity. If this is the case, signals of dark matter scatterings with nuclei or with electrons could be observed in a dedicated detector at Earth, not only establishing the particle nature of the dark matter, but also opening the possibility of studying the characteristics of the dark sector. 
 
 A crucial ingredient in the calculation of the interaction rate is the dark matter flux at the location of the detector. The flux depends on the number density of dark matter particles at the Solar System, which in turn depends on the mass density and the dark matter mass, as well as on the velocity distribution of dark matter particles. None of these quantities are positively known. It is common in the literature to adopt the Standard Halo Model (SHM), with a local mass density $\sim 0.3\,{\rm GeV}/{\rm cm}^3$, based on extrapolations of astronomical observations at kpc scales to the small scales of our Solar System, as well as an isotropic Maxwell-Boltzmann velocity distribution, based on theoretical considerations. On the other hand, these considerations are known to be inaccurate. In fact, numerical $N$-body simulations suggest a velocity distribution at the Solar System which is quantitatively different to the isotropic Maxwell-Boltzmann distribution, although having qualitatively a similar form \cite{Vogelsberger:2008qb, Sloane:2016kyi,Evans:2000gr,Evans:2018bqy}.
 
 The theoretical modeling of the dark matter phase space distribution in the Solar System normally assumes that the Milky Way is an isolated galaxy. However, the Milky Way is one among the various members of the Local Group, which include M31, M33 and several dwarf galaxies. Various astronomical observations suggest that the Local Group contains a diffuse component of dark matter, not belonging to the isolated halos of its subsystems, and distributed roughly homogeneously over the cluster \cite{1959ApJ...130..705K, 2008gady.book.....B, Cox:2007nt}.
 This component would permeate the Solar System, and would contain particles moving with speeds larger than the escape velocity from the Milky Way. Further, it has been suggested that the Virgo Supercluster could also contain a diffuse component \cite{Makarov:2010di,Karachentsev_2018}. The true dark matter flux, which includes the non-galactic components, would therefore be qualitatively different to the one expected from the Standard Halo Model. 
 
 In this paper we will investigate the impact of  non-galactic dark matter in direct detection experiments. Related analyses have been presented in  \cite{Baushev:2012dm, Freese:2001hk,Green:2000ga,Stiff:2001dq}, focusing on nuclear recoils induced by dark matter particles with mass in the electroweak range. Here we concentrate instead on light dark matter particles, for which modifications in the high velocity part of the flux are expected to have a more significant impact in direct detection experiments, and consider specifically the impact of the non-galactic dark matter from the Local Group and the Virgo Supercluster diffuse components. The paper is organized as follows. In Section \ref{sec:flux} we recapitulate the various components of the local dark matter flux, in Sections \ref{sec:nuclear_recoils} and \ref{sec:electron_recoils} we investigate the impact of non-galactic dark matter in nuclear and electron recoils, respectively,  and in section \ref{sec:conclusions} we present our conclusions. We also include an appendix describing some technical aspects of the derivation of the limits from direct detection experiments. 

\section{Dark matter flux at the Solar System}
\label{sec:flux}

The signal rate at a direct detection experiment crucially depends on the dark matter flux at the detector. One can identify various possible contributions to the local dark matter flux. The most likely contribution stems from dark matter particles of the Milky Way halo. The dark matter density at the location of the Solar System and their velocity distribution is uncertain. It is common in the literature to adopt the values of the Standard Halo Model, namely a local density $\rho^{\rm loc}_{\rm SHM}=0.3\,{\rm GeV}/{\rm cm}^{3}$  and a velocity distribution (expressed in the Galactic frame):
\begin{align}
\label{eq:f_MB}
f_\mathrm{SHM} ( \vec{ v } ) = \frac { 1} { ( 2\pi \sigma _ { v } ^ { 2} ) ^ { 3/ 2} N _ { \text{esc} } } \exp \left[ - \frac { v^2} { 2\sigma_v^2 } \right]\quad \text{for } v \leq v_{\rm esc}\,,
\end{align}
where, $v=|\vec v|$, $\sigma_v = 156$ km/s is  the velocity dispersion \cite{Kerr:1986hz,Green:2011bv}, and $v_\mathrm{esc}=544$ km/s is the escape velocity from our Galaxy~\cite{Smith:2006ym,Piffl:2013mla}  (for a recent study, see  \cite{Deason_2019}).
Further, $N _ { \text{esc} }$ is a normalization constant, given by:
\begin{align}
N _ { \text{esc} } = \operatorname{erf} \left( \frac { v _ { \text{esc} } } { \sqrt { 2} \sigma _ { v } } \right) - \sqrt { \frac { 2} { \pi } } \frac { v _ { \text{esc} } } { \sigma _ { v } } \exp \left( - \frac { v _ {\mathrm{esc}} ^ { 2} } { 2\sigma _ { v } ^ { 2} } \right)\,.
\end{align}
For our chosen parameters, $N_{\rm esc}\simeq 0.993$.
The contribution to the local dark matter flux from the Milky Way halo then reads:
\begin{align}
{\mathscr F}_{\rm SHM}(\vec v)=  \frac{\rho_{\rm SHM}^{\rm loc}}{m_{\rm DM}} v f_{\rm SHM}(\vec v)\;,
\end{align}
with  $m_{\rm DM}$ the dark matter mass. It should be noted that the true local density and velocity distribution may differ from these commonly adopted values \cite{Read:2014qva,Green:2017odb}, and a number of refinements to the SHM have been proposed in recent years \cite{Evans:2018bqy, Bozorgnia:2017brl}. The impact of these deviations on direct detection experiments has been discussed in various works \cite{Feldstein:2014ufa,Ibarra:2018yxq,Fowlie:2018svr}. Moreover, the Milky Way halo could contain substructures, such as streams or subhalos, which may also enhance the dark matter flux at the location at the Solar System \cite{OHare:2018trr}. The probability of a sizable enhancement is, on the other hand, modest \cite{Ibarra:2019jac}.

Dark matter particles from the Milky Way halo and its substructures are expected to be gravitationally bound to our Galaxy, and therefore to have speeds at the Solar System smaller than $v_{\rm esc}$. On the other hand, there could be a contribution to the local flux from non-galactic dark matter, not gravitationally bound to the Milky Way, and with larger speeds. This component would dominate the high-velocity tail of the dark matter flux, and could impact significantly the scattering rate of light dark matter particles in a detector. In this work we consider the contribution from the Local Group and from the Virgo Supercluster diffuse components.

The Local Group consists of two very massive galaxies, Milky Way and Andromeda, the less massive Triangulum galaxy, and a number of dwarf galaxies. The Local Group could contain dark matter particles bound to the full system, but not to the individual galaxies  \cite{1959ApJ...130..705K, Cox:2007nt, 2005ESASP.576..651W, 2008gady.book.....B,Karachentsev_2018}. The dark matter envelope of the Local Group was modeled in  \cite{Baushev:2012dm}. At scales $\sim$ 300-600 kpc away from the Milky Way center, approximately corresponding to the size of the Milky Way Roche lobe in the system Milky Way-M31, this envelope amounts to a population of dark matter particles contributing to the local density of dark matter at the Solar System with $\rho_{\rm LG} \sim 10^{-2}$ GeV/cm$^{3}$, and moving isotropically with speed  $v_{\rm LG}\sim 600$ km/s, which is roughly the escape velocity of the system Milky Way-envelope at the position of the Solar System. These particles present a narrow velocity distribution, $\sigma_{v.{\rm LG}} \sim 20$ km/s, due to the small difference in the gravitational potential of the system at the boundaries of the envelope (from $\sim$ 300 to $\sim$ 600 kpc to the position of the Solar System). The contribution from the Local Group to the dark matter flux at the location of the Solar System can then be written as:
\begin{align}
{\mathscr F}_{\rm LG}(\vec v)=  \frac{\rho_{\rm LG}^{\rm loc}}{m_{\rm DM}} v \delta^{3}(\vec{v} - \vec{v}_{\rm LG}).
\end{align}

Dark matter in the Virgo Supercluster could also contribute to the dark matter flux in the Solar System. Measurements estimate the average density of the diffuse component to be $\sim 10^{-6}$ GeV/cm$^{3}$ \cite{Makarov:2010di}. However, the gravitational focusing due to the Local Group leads to a density at the location of the Sun enhanced by a factor $\sim 1+\frac{v_{\rm esc}^{2}}{v_{\rm VS}^{2}}$, which results into a contribution to the total local density of $\rho_{\rm VG}^{\rm loc}\sim 10^{-5}$ GeV/cm$^{3}$.  Current knowledge on the dark matter velocity distribution in the Virgo Supercluster is much poorer. Following \cite{Baushev:2012dm}, we assume that the dark matter particles have a velocity dispersion comparable to that of the observable members of the Virgo Supercluster, which yields speeds for the Virgo Supercluster dark matter particles at Earth to be (at least) $v_{\rm VS}\sim 1000$ km/s. The contribution from the  Virgo Supercluster to the dark matter flux at the location of the Solar System can be written as:
\begin{align}
{\mathscr F}_{\rm VS}(\vec v)=  \frac{\rho_{\rm VS}^{\rm loc}}{m_{\rm DM}} v \delta^{3}(\vec{v} - \vec{v}_{\rm VS}).
\end{align}

We then model the dark matter flux at the position of the Solar System as the normalized sum of these various contributions:
\begin{align}
{\mathscr F}(\vec v)={\mathscr F}_{\rm SHM}(\vec v)+{\mathscr F}_{\rm  LG}(\vec v)+{\mathscr F}_{\rm \rm VS}(\vec v),
\label{eq:total_flux}
\end{align}
where we adopt values for the local density of each component derived in \cite{Baushev:2012dm}, such that the total sum yields the canonical value of the local density $\rho^{\rm loc}=0.3$ GeV/cm$^{3}$ used by  direct detection experiments: $\rho^{\rm loc}_{\rm SHM}=0.26$ GeV/cm$^{3}$ ($\sim 88 \%$), $\rho^{\rm loc}_{\rm LG}=0.037$ GeV/cm$^{3}$ ($\sim 12 \%$), and $\rho^{\rm loc}_{\rm VS}=10^{-5}$ GeV/cm$^{3}$ ($\sim 0.00003 \%$).

The parameters of the non-galactic flux components are subject to uncertainties, {\it e.g.} the determination of the mass of the Local Group envelope \cite{Cox:2007nt}. The values we adopt in this work can be regarded as conservative, and are meant to illustrate the potential sensitivity of non-galactic dark matter in light dark matter searches.  Additional investigations about the dynamics of the Virgo Supercluster and Local Group members ({\it e.g.} \cite{van_der_Marel_2019, Kourkchi_2017,Shaya_2017}), in combination with a more refined modeling of the dark matter distribution in these objects, will be pivotal to better determine the phase-space distribution of non-galactic dark matter at the Solar System.

\section{Impact on nuclear recoils: CRESST III and XENON1T}
\label{sec:nuclear_recoils}

The differential rate of nuclear recoils induced by scatterings of dark matter particles traversing a detector at the Earth is given by \cite{Lewin:1995rx,Cerdeno:2010jj}:
\begin{align}
\frac{dR}{dE_R}= \sum_i \frac{ \xi_i}{m_{A_i} } \int_{v \geq v^i_{\rm min}(E_R)} \text{d}^3 v {\mathscr F}( \vec{v} + \vec v_\odot)\, \frac{\text{d}\sigma_i}{\text{d}E_R}(v, E_R) \,.
\label{eq:diff_scattering_rate}
\end{align}
Here, $\vec v$ is the dark matter velocity in the rest frame of the detector, ${\mathscr F}( \vec{v}+\vec v_\odot)$ is the dark matter flux in the galactic frame, given in Eq.(\ref{eq:total_flux}), and $\vec v_\odot$ is the velocity of the Sun with respect to the Galactic frame,  given by $\vec{v}_{\odot}=\vec{v}_{\text{LSR}}+\vec{v}_{\odot,{\rm  pec}}$, where $\vec{v}_{\text{LSR}}\approx (0,220,0)$ km/s is the velocity of the local standard of rest (LSR) and $\vec{v}_{\odot,{\rm pec}}=(11.1,12.24,7.25)$ km/s is the
peculiar velocity of the Sun with respect to the LSR~ \cite{McCabe:2013kea}. Further, $v^i_{\rm min}(E_R) = \sqrt{m_{A_i} E_R/(2 \mu_{A_i}^2)}$ is the minimal speed necessary for a dark particle to induce a recoil with energy $E_R$ of the nucleus $i$ with mass $m_{A_i}$ and mass fraction $\xi_{i}$ in the detector.  $\text{d}\sigma_i/\text{d}E_R$ is the differential cross section for the elastic scattering of dark matter off the nucleus $i$ producing a nuclear recoil energy $E_R$. For spin-independent interactions, the differential dark matter-nucleus cross section reads,
\begin{align}\label{eq:sidiffcross}
    \frac{d\sigma^{\rm SI}_i}{dE_{R}}(v,E_R)=\frac{m_{A_{i}}}{2\mu_{\mathscr N}^{2}v^{2}}A_i^{2}\sigma_{\mathscr N}^{\rm SI}F_i^{2}(E_{R})\;,
\end{align}
where we have assumed for simplicity a Majorana dark matter candidate that couples with equal strength to protons and neutrons. Here $A_i$ is the mass number of the nucleus $i$,  $F_i^{2}(E_{R})$ is the nuclear form-factor, for which we adopt the Helm prescription, and $\sigma_{\mathscr N}^{\rm SI}$ is the spin-independent scattering cross section off the nucleon ${\mathscr N} =n,p$ at zero momentum transfer, which depends on the details of the dark matter model.
For spin-dependent interactions, the differential dark matter-nucleus cross section reads
\begin{align}
    \frac{d\sigma^{\rm SD}_i}{dE_{R}}(v, E_R)=\frac{2\pi m_{A_i} }{3\,\mu_{\mathscr N} \,v^2 \,(2J+1)}  \sigma^{\rm SD}_{\mathscr N}S_{A_i}(E_R)\;,
\end{align}
with  $J$  the total spin of the nucleus, $S_{A_i}(E_R)$ the nuclear structure function, and $\sigma_{\mathscr N}^{\rm SD}$ is the spin-dependent scattering cross section off the nucleon ${\mathscr N} =n,p$ at zero momentum transfer. 

The total recoil rate can be calculated from the differential rate using:
\begin{align}
R=  \int_{0}^\infty \text{d}E_R \, \epsilon_i (E_R) \frac{dR}{dE_R},
\label{eq:scattering_rate}
\end{align}
with  $\epsilon_i(E_R)$ the efficiency, defined as the probability to detect the recoil of the target nucleus $i$ with energy $E_R$. Finally, the total number of expected recoil events at a direct detection experiment reads ${\cal N}=R\cdot \mathcal{E}$, with $\mathcal{E}$ the exposure (i.e.~mass multiplied by live-time) of the experiment.

\begin{figure}[t]
\includegraphics[width=0.45\textwidth]{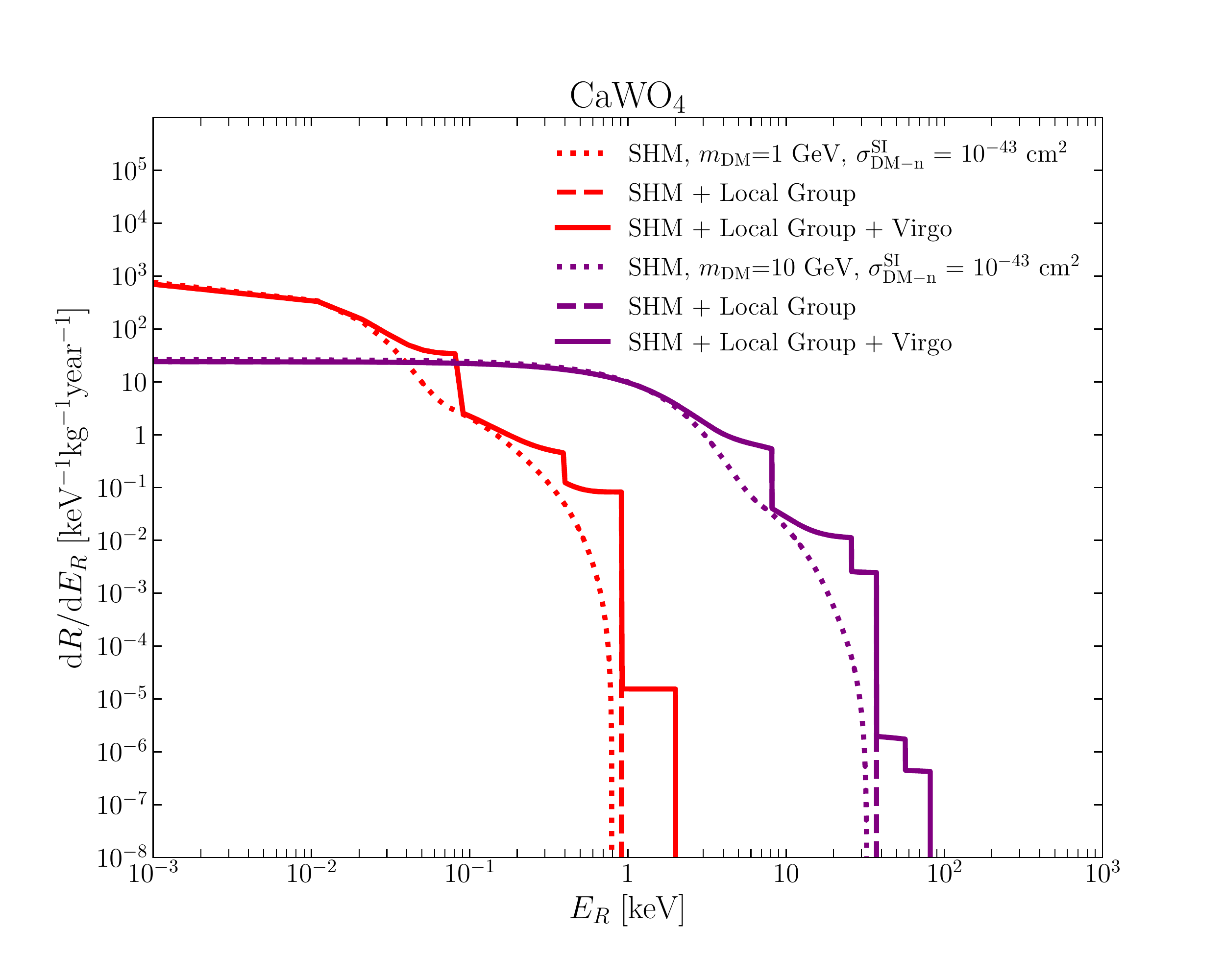}
\hspace{0.01\textwidth}
\includegraphics[width=0.45\textwidth]{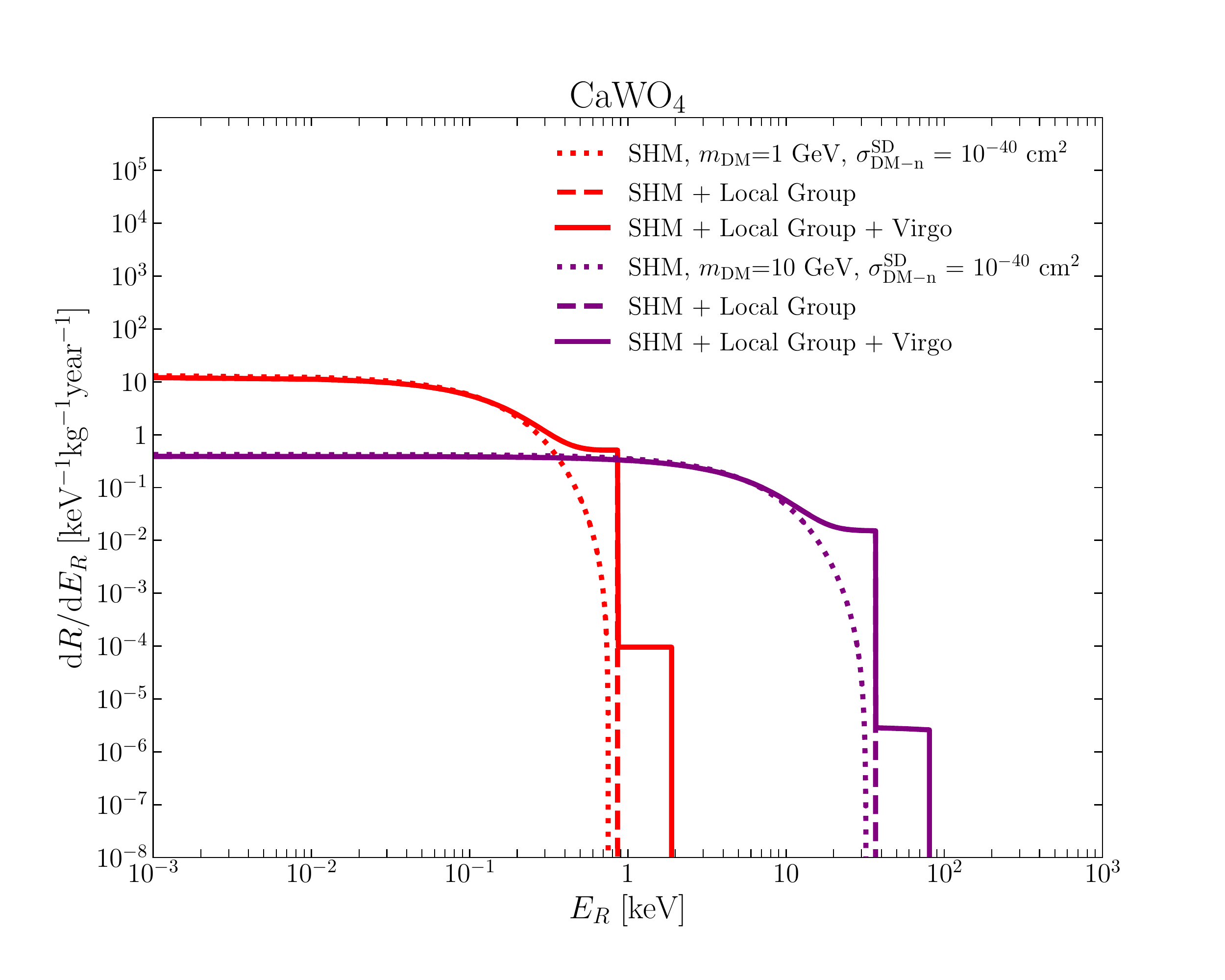}\\
\includegraphics[width=0.45\textwidth]{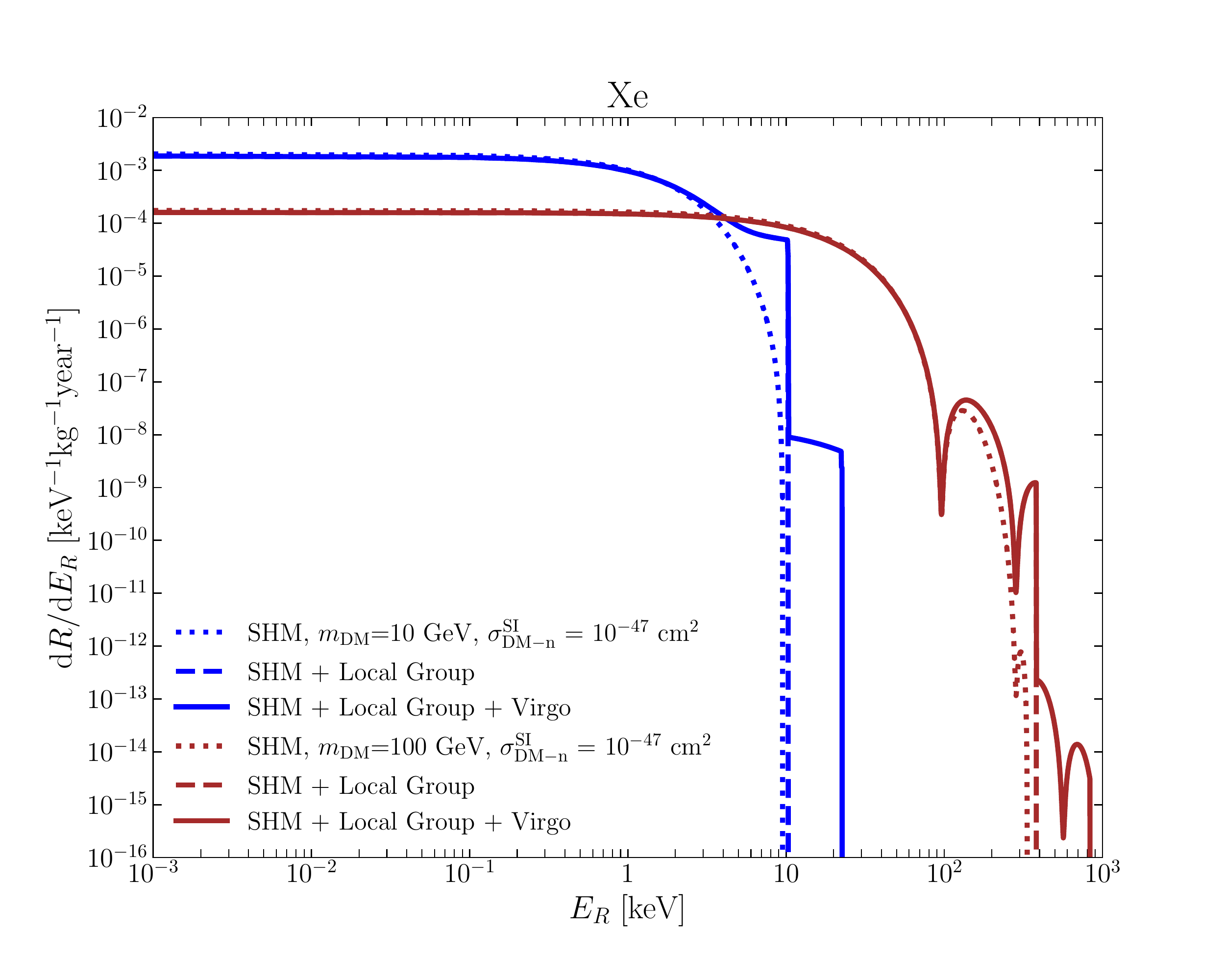}
\hspace{0.01\textwidth}
\includegraphics[width=0.45\textwidth]{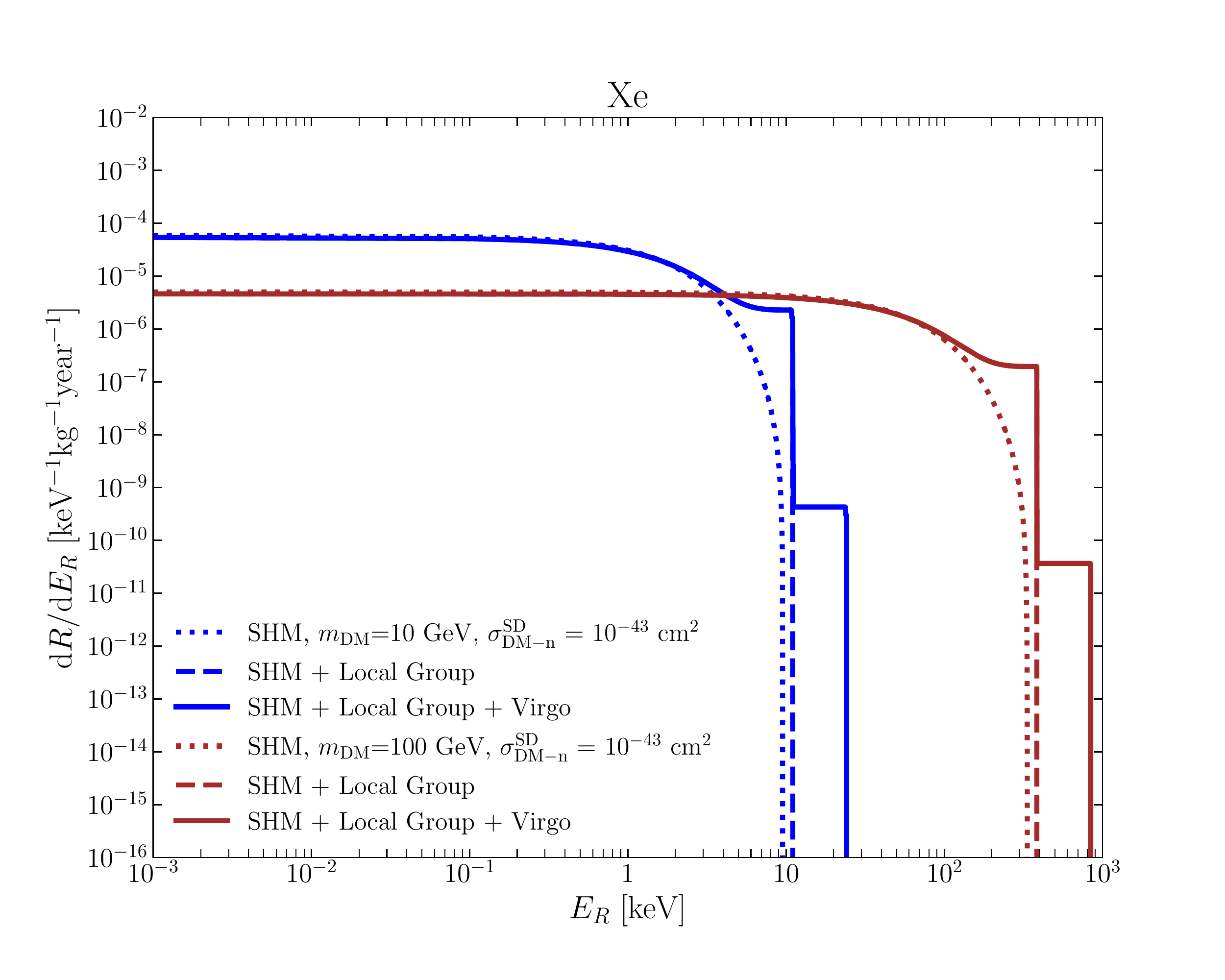}
\caption{\small Dark matter-nucleon differential recoil rate with a CaWO$_4$ target (top panels) and a Xe target (bottom panels), assuming  spin-independent (left panels) or spin-dependent scattering (right panels), for two exemplary choices of the dark matter mass, and for values of the cross-section close to the current upper limit from experiments. The dotted lines indicate the differential rate expected from the Standard Halo Model, the dashed lines include also the contribution from the Local Group, and the solid lines also the contribution from the Virgo Supercluster. }
\label{fig:rate_sisd}
\end{figure}

We show in Figure \ref{fig:rate_sisd} the different contributions of the dark matter flux to the differential recoil spectrum with a CaWO$_4$ target (top panels) and a Xe target (bottom panels), assuming a spin-independent interaction only (left  panel) or the spin-dependent interaction only (right panel), considering for each case values of the cross section near the current sensitivity of experiments. We take two exemplary choices for the dark matter mass: $m_{\rm DM}=1$ GeV (red) and 10 GeV (purple) for the CaWO$_4$ target, while $m_{\rm DM}=10$ GeV (blue) and 100 GeV (brown) for the Xe target. For each mass, we show the differential recoil rate induced just by the SHM component (dotted lines), as well as the differential rate including the contribution from the Local Group (dashed lines) and including also the Virgo Supercluster (solid lines).

The new contributions modify the differential rate at the highest recoil energies, causing deviations from the differential recoil spectrum expected within the SHM. The non-galactic contributions could increase the total number of detected recoil events, thus increasing the discovery potential of direct detection experiments. Further, due to the small velocity dispersion of the non-galactic dark matter particles, the differential recoil spectrum presents step-like features which could be crucial to distinguish a dark matter signal from irreducible backgrounds with a smoother spectrum, such as the one arising from the coherent elastic scattering of solar neutrinos off nuclei~\cite{Drukier:1983gj,Billard:2013qya}.

Concretely, the ratio between the contributions to the differential recoil rate from the Local Group, and the contribution from the Standard Halo Model can be estimated analytically. Assuming a single target, it reads:
\begin{align}
\frac{dR/dE\Big|_{\rm LG}}{dR/dE\Big|_{\rm SHM}} \approx  \frac{\rho_{\rm LG}^{\rm loc}}{\rho_{\rm SHM}^{\rm loc}}
\frac{\sigma_v}{v^*_{\rm LG}} \sqrt{\frac{\pi}{2}}N_{\rm esc} 
\frac{\theta\Big(v^*_{\rm LG}-v_{\rm min}(E_R)\Big)}{e^{-\frac{v_{\rm min}^2(E_R)}{2\sigma_v^2}}-
e^{-\frac{v_{\rm esc}^{*2}}{2\sigma_v^2}}}\;,
\label{eq:ratio_diff_rates}
\end{align}
where starred variables refer to the detector frame: $\vec v^*=\vec v+\vec v_\odot$.
For most values of $E_R$ the Local Group contributes negligibly, due to the large suppression factor $(\rho_{\rm LG}^{\rm loc}/\rho_{\rm SHM}^{\rm loc} )(\sigma_v/v^*_{\rm LG})$. However, for large values of $E_R$, the Local Group can provide a comparable or even dominant contribution to the rate. Concretely, the Local Group provides a contribution to the differential rate comparable to the one from the Standard Halo Model when 
\begin{align}
E_R\gtrsim E_R^{\rm max,SHM}\Big[1- \frac{2 \sigma_{v}^{2}}{v_{\rm esc}^{*2}}\rm ln\Big(\frac{\rho_{\rm LG}^{\rm loc}}{\rho_{\rm SHM}^{\ loc}}
\frac{\sigma_v}{v^*_{\rm LG}}   \sqrt{\frac{ \pi}{2}}N_{\rm esc}e^{\frac{v_{\rm esc}^{*2}}{2\sigma_v^2}}+1\Big)\Big]\;,
\end{align}
with  $E_R^{\rm max, SHM}=2\mu_{A}^2 v_{\rm esc}^{*2}/m_{A}$ the largest possible recoil energy within the SHM, and dominates the recoil spectrum up to $E_R^{\rm max,LG}=2\mu_{A}^2 v_{\rm LG}^2/m_{A}= v_{\rm LG}^{*2}/v_{\rm esc}^{*2} E_R^{\rm max,SHM}$, which is the largest possible recoil energy from DM particles in the Local Group envelope.  For our adopted values, one obtains $E_R\gtrsim 0.29 E_R^{\rm max,SHM}$ and extends up to $E_R^{\rm max,LG}\simeq 1.2 E_R^{\rm max,SHM}$; these numbers are in qualitative agreement with Fig.~\ref{fig:rate_sisd}. Analogous expressions hold for the contribution from the Virgo Supercluster: one obtains that this contribution dominates over the SHM one for $E_R\gtrsim 0.94 E_R^{\rm max,SHM}$ and extends up to $E_R^{\rm max,VS}\simeq 2.5 E_R^{\rm max,SHM}$.

\begin{figure}[t!]
\includegraphics[width=0.45\textwidth]{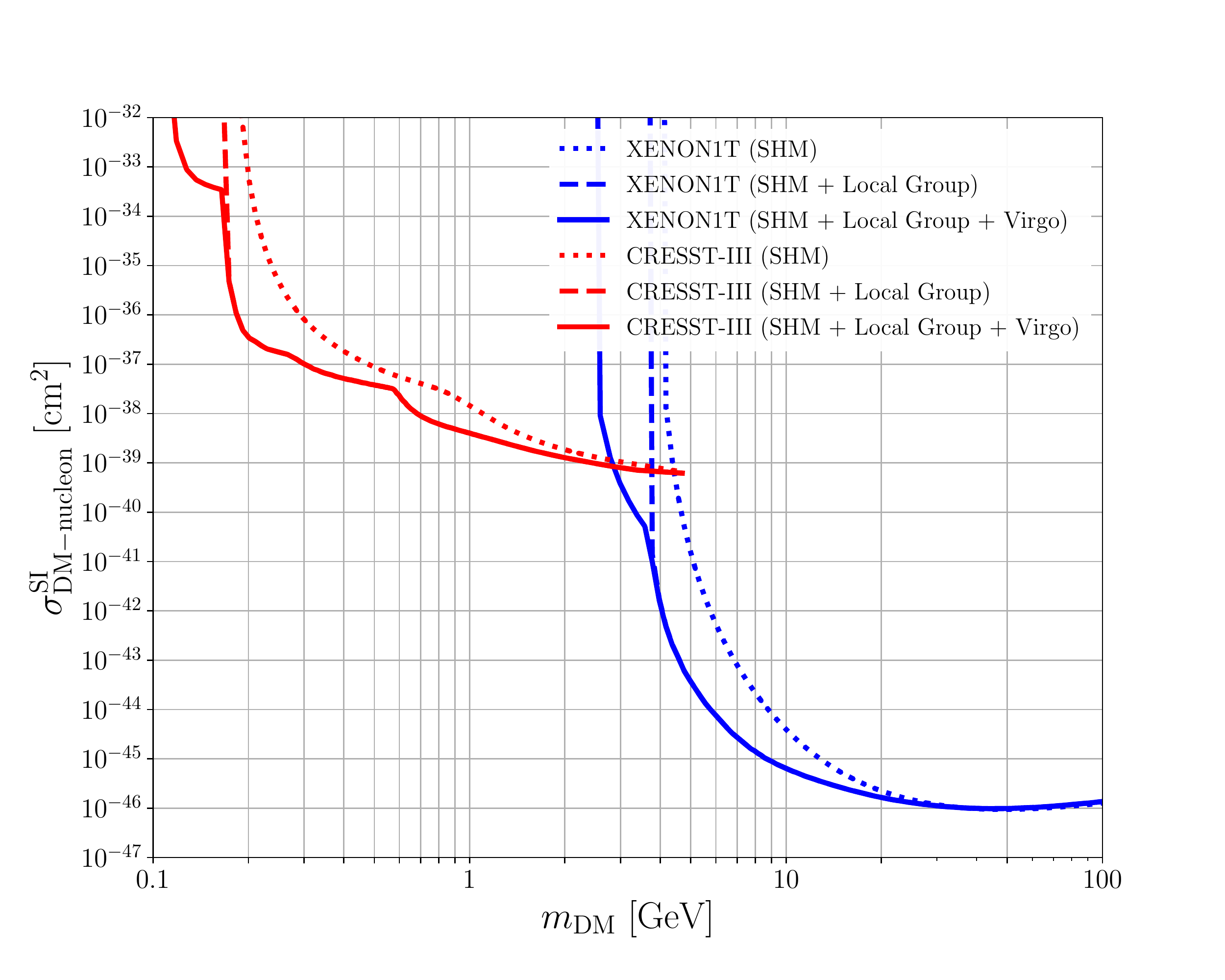}
\hspace{0.05\textwidth}
\includegraphics[width=0.45\textwidth]{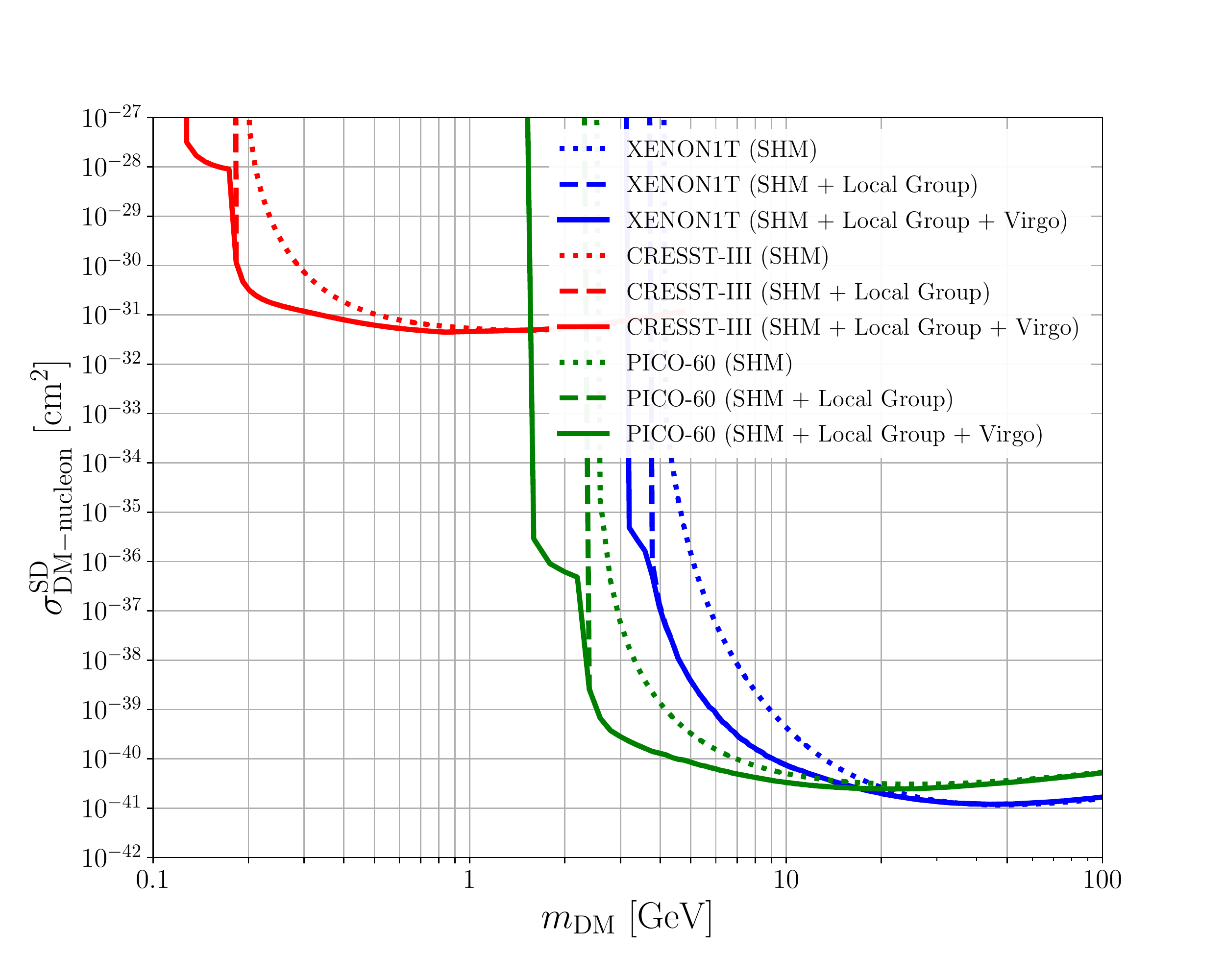}
\caption{\small Upper limits at the 90$\%$ C.L. on the spin-independent (left panel) and spin-dependent (right panel) dark matter-nucleon cross-section from the null search results from the XENON1T (blue), CRESST-III (red) and PICO-60 (green) experiments, assuming equal coupling to protons and neutrons.
The dotted line indicates the upper limit derived under the assumption that only galactic dark matter, described by the Standard Halo Model, contributes to the dark matter flux at the Solar System. The dashed lines show the impact of including in the flux also the  non-galactic dark matter component from the Local Group and the solid lines show the impact of including also the diffuse component of the Virgo Supercluster.}
\label{fig:sisd}
\end{figure}

We show in Figure \ref{fig:sisd} the upper limits on the dark matter-nucleon spin independent (left panel) or spin-dependent (right panel) scattering cross section as a function of the dark matter mass from the non-observation of dark matter induced nuclear recoils at the CRESST-III or the XENON1T experiments. The potential impact of the dark matter envelope of the Local Group for the search of light dark matter is apparent from  the Figure. For $m_{\rm DM}\lesssim 1$ GeV, this contribution can enhance the recoil rate at the CRESST experiment by at least a factor $\sim 2$. As the dark matter mass decreases, the enhancement becomes more and more important, and even allows to probe masses for which the galactic dark matter would not induce detectable recoils. Similarly, for $m_{\rm DM}\lesssim 10$ GeV, the recoil rate at the XENON1T experiment is increased by at least a factor $\sim 10$, thus increasing the discovery potential of the experiment.~\footnote{For very large cross-sections, the dark matter flux could be attenuated in its passage through the Earth before reaching the detector~\cite{Armengaud:2019kfj, Kouvaris:2015laa}. We estimate this attenuation to be negligible for the values shown in the Figure.}

\section{Impact on electron recoils: XENON10 and XENON100}
\label{sec:electron_recoils}

The dark matter-electron scattering rate in liquid xenon has been discussed at length in the literature \cite{Essig:2012yx,Essig:2017kqs,Catena:2019gfa}. The differential ionization rate reads:
\begin{align}
\frac{dR_{ion}}{d\text{ln}E_{er}}= N_{T}\sum_{n,l}  \int_{v \geq v_{\rm min}^{nl}(E_{er})} \text{d}^{3} v {\mathscr F}(\vec{v}+\vec v_\odot) \, \frac{\text{d} \sigma_{\rm ion}^{nl}}{d\text{ln}E_{er}}(v, E_{er}) \,,
\label{eq:diff_ionization_rate}
\end{align}
where $N_{T}$ is the number of target nuclei and
\begin{align}
v_{\rm min}^{nl}(E_{er}) =\sqrt{\frac{2}{m_{\rm DM}}(E_{er}+|E^{nl}|)}
\label{eq:vmin_e}
\end{align}
is the minimum dark matter velocity necessary to ionize a bound electron in the $(n,l)$ shell of a xenon atom (with energy  $E^{nl}$), giving a free electron with energy $E_{er}$. Further, $d\sigma_{\rm ion}^{nl} /d\text{ln}E_{er}$ is the differential ionization cross section, given by: 
\begin{align}
    \frac{d\sigma_{ion}^{nl }}{d\text{ln}  E_{er}}(v,E_{er})=\frac{\bar{\sigma}_{\rm DM-e}}{8 \mu_{\rm DM, e}^{2}v^{2}}\int_{q^{nl}_{\rm min}}^{q^{nl}_{\rm max}} dq q \left |f_{ion}^{nl}(k',q)  \right |^{2} \left |F_{\mathrm{DM}}(q)  \right |^{2}.
\end{align}
Here, $\mu_{\rm DM,e}$ is the reduced mass of the dark matter-electron system, 
$\bar{\sigma}_{\rm DM-e}$ is the dark matter-free electron scattering cross section at fixed momentum transfer $q=\alpha m_{e}$, $\left|f_{ion}^{nl}(k',q)  \right |^{2}$ is the ionization form factor of an electron in the $(n,l)$ shell with final momentum $k'=\sqrt{2m_{e}E_{er}}$ and momentum transfer $q$, and $F_{\mathrm{DM}}(q)$ is a form factor that encodes the $q$-dependence of the squared matrix element for dark matter-electron scattering and depends of the mediator under consideration. Note that the momentum transfer is not univocally determined, due to the fact that the electron momentum in the atomic orbital is not fixed.
The maximum and minimum values of the momentum transfer producing an electron recoil with energy $E_{er}$ from the interaction of a dark matter particle with speed $v$ with a bound electron in the $(n,l)$ shell are:
\begin{align}
    q^{nl}_{\substack{{\rm max}\\{\rm min}}}(E_{er})= m_{\rm DM} v \left[1\pm\sqrt{1-\left(\frac{v_{\rm min}^{nl}(E_{er})}{v}\right)^2}\right],
    \label{eq:q_max_min}
\end{align}
with $v_{\rm min}^{nl}(E_{er})$ defined in Eq.~(\ref{eq:vmin_e}). 
Finally, the total number of expected ionization events reads ${\cal N}=R_{ion}\cdot \mathcal{E}$,  with 
$R_{ion}$ the total ionization rate, calculated from integrating Eq.(\ref{eq:diff_ionization_rate}) over all possible recoil energies, and 
$\mathcal{E}$ the exposure ({\it i.e.} mass multiplied by live-time) of the experiment.~\footnote{The efficiency function of XENON10 and XENON100 is taken into account when calculating the upper limit on the number of signal events, as described in Appendix \ref{sec:details}.} In our analysis, we consider the ionization of electrons in the three outermost orbitals (with binding energies in eV shown in parenthesis): $5p^{6}$ (12.4), $5s^{2}$ (25.7) and $4d^{10}$ (75.6). The corresponding ionization form factors were calculated using the software \texttt{DarkARC} \cite{Catena:2019gfa}. For the dark matter form factor, we adopt two different parametrizations: the case of a heavy hidden photon $A'$ mediator $m_{A'} \gg q$, with $F_{\mathrm{DM}}(q)$=1, and an ultralight hidden photon $m_{A'} \ll q$, with $F_{\mathrm{DM}}(q)=\alpha^{2}m_{e}^{2}/q^{2}$.

\begin{figure}[t!]
\includegraphics[width=0.45\textwidth]{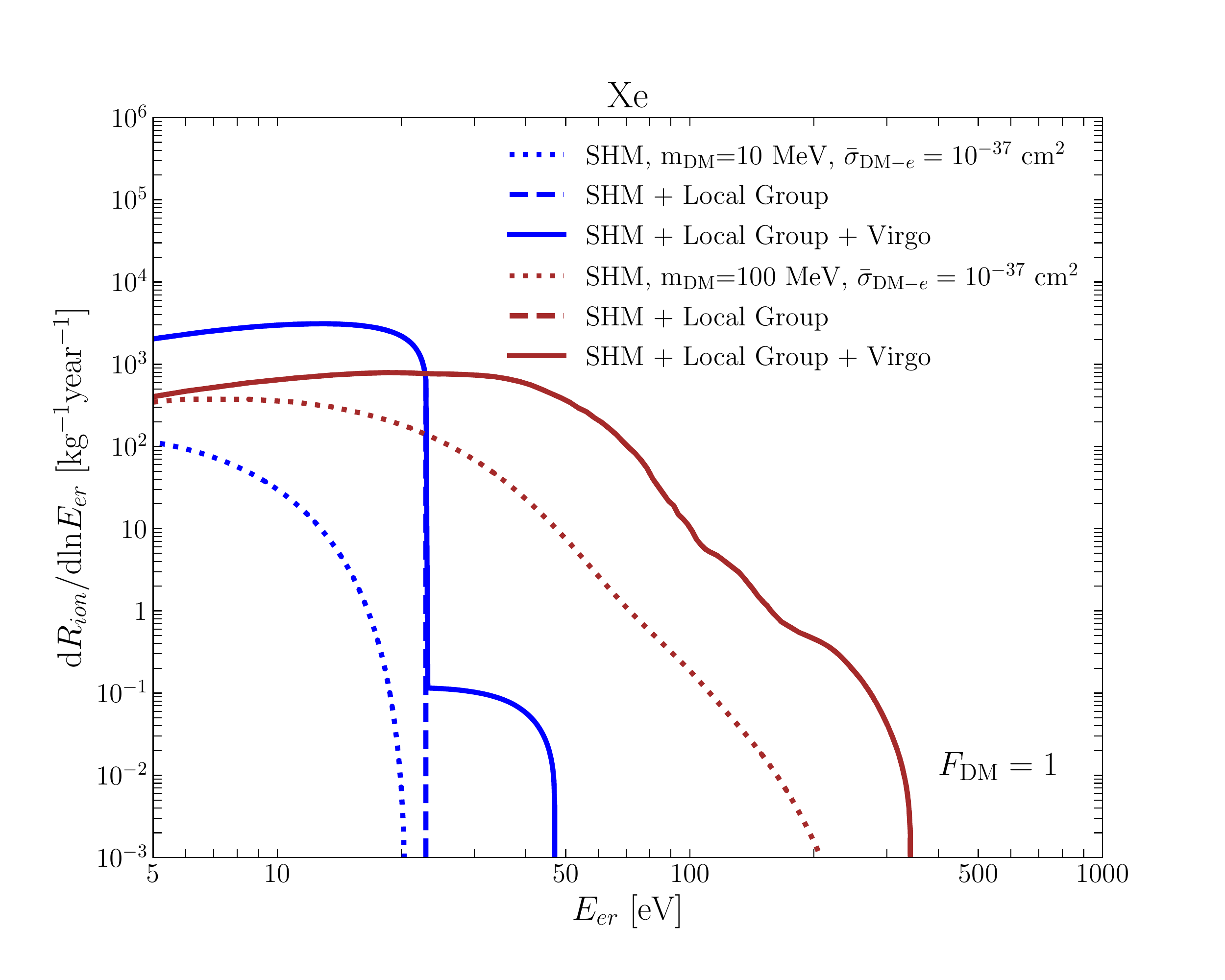}
\hspace{0.05\textwidth}
\includegraphics[width=0.45\textwidth]{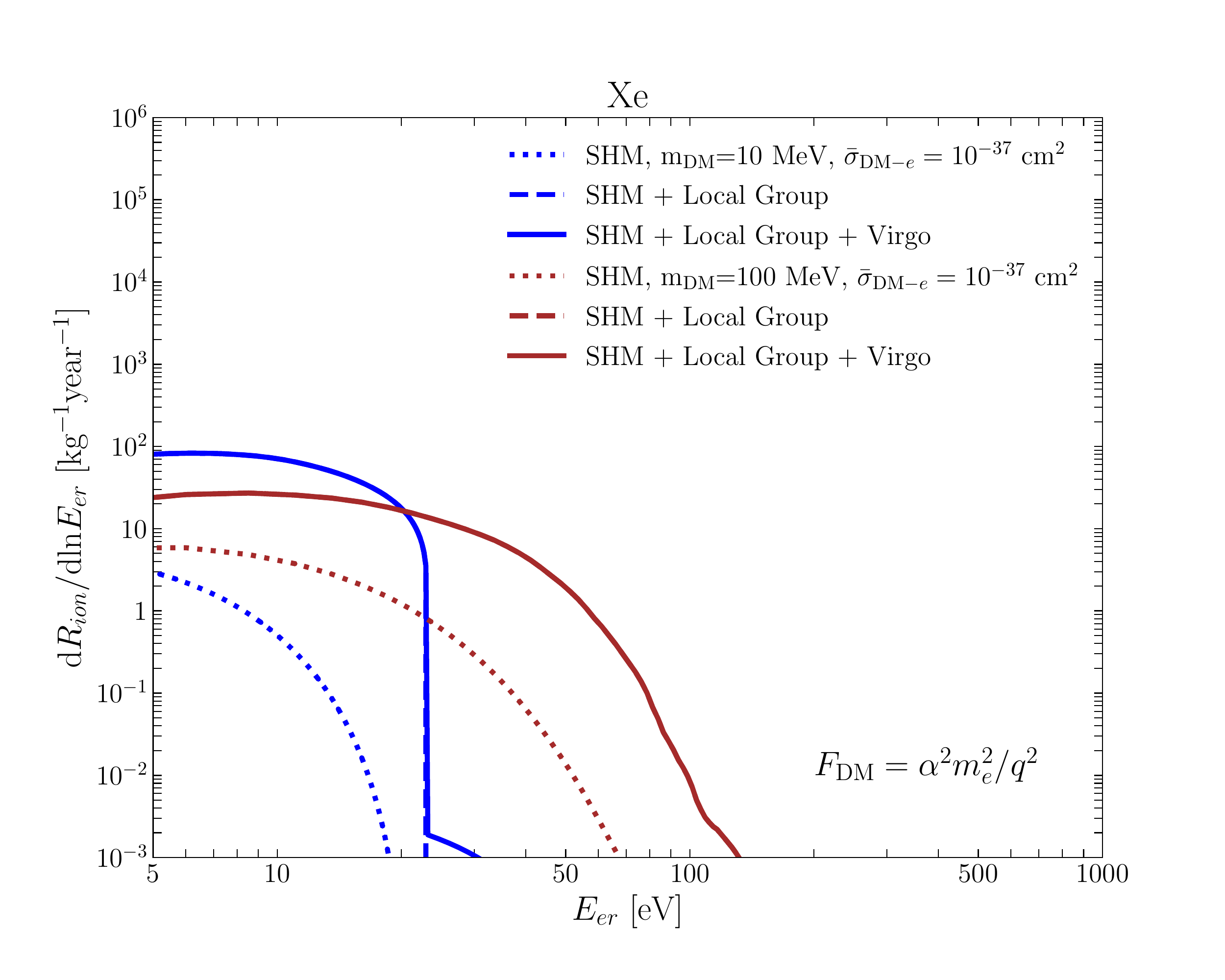}
\caption{\small Dark matter-electron differential recoil rate with a Xe target, assuming  an interaction mediated by a heavy hidden photon (left panel) or an ultralight hidden photon (right panel), for two exemplary choices of the dark matter mass: $m_{\rm DM}=$100 MeV (brown) and $m_{\rm DM}=$10 MeV (blue), and values of the cross-section close to the current upper limit from experiments.
The dotted lines indicate the differential rate expected from the Standard Halo Model, the dashed lines include also the contribution from the Local Group, and the solid lines also the contribution from the Virgo Supercluster. }
\label{fig:rate_e}
\end{figure}

We show in Figure \ref{fig:rate_e} the different contributions to the dark matter-electron differential recoil rates at an experiment with a Xe target, for two exemplary choices of the dark matter mass, 10 MeV and 100 MeV, and for the aforementioned two parametrizations. For each case, we assume a value of the cross section near the current sensitivity of experiments.  Further, for each mass, we show the differential recoil rate induced just by the SHM component (dotted lines), as well as the enhancement in the differential recoil rate induced by dark matter from the Local Group (dashed lines) and by both the Local Group and the Virgo Supercluster (solid lines).

As can be seen in the plot, the non-galactic dark matter can have a considerable impact on the electron recoil spectrum. Similarly to the nuclear recoils, the Local Group provides a contribution to the rate $\propto (\rho_{\rm LG}^{\rm loc}/\rho_{\rm SHM}^{\rm loc})( \sigma_v/v^*_{\rm LG}$), but due to the values of the form factors, the enhancement is numerically larger. More importantly, the impact of the Local Group contribution is significant over a wider range of recoil energies, and not just close to the kinematical threshold, which is due to the fact that the momentum transfer is not fixed for scatterings off electrons in an atomic orbital. For very small mass, such as for $m_{\rm DM}=10$ MeV, the momentum transfer is not fixed, but takes values within a small range, {\it cf.} Eq.~(\ref{eq:q_max_min}). Therefore the contribution to the recoil spectrum from the Local Group resembles a step function (as for nuclear recoils). For larger masses, the contribution from the Local Group to the recoil spectrum is the superposition of various step functions (corresponding to different values of the momentum transfer), generating a featureless spectrum that extends to lower recoil energies.

\begin{figure}[t!]
\includegraphics[width=0.45\textwidth]{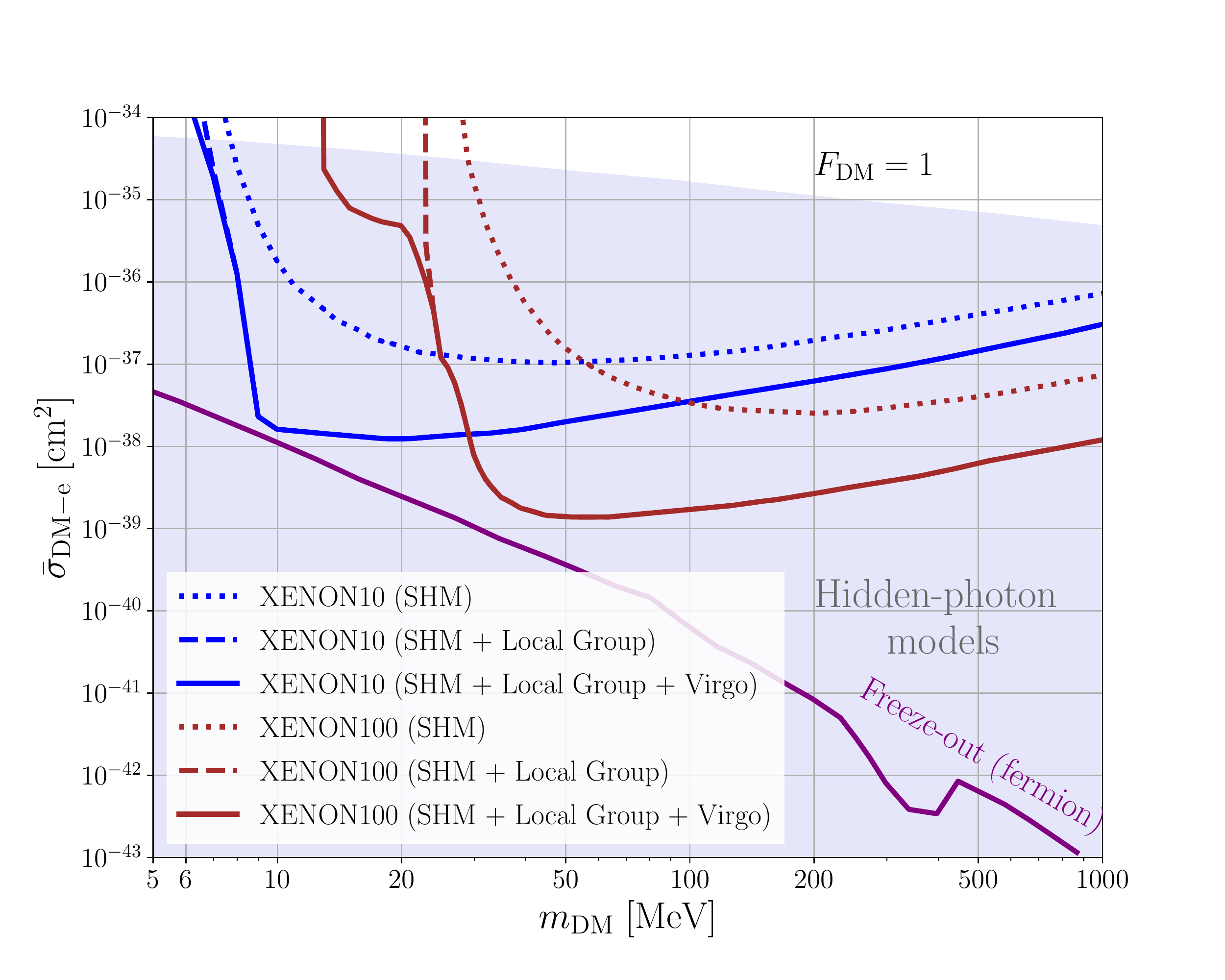}
\hspace{0.05\textwidth}
\includegraphics[width=0.45\textwidth]{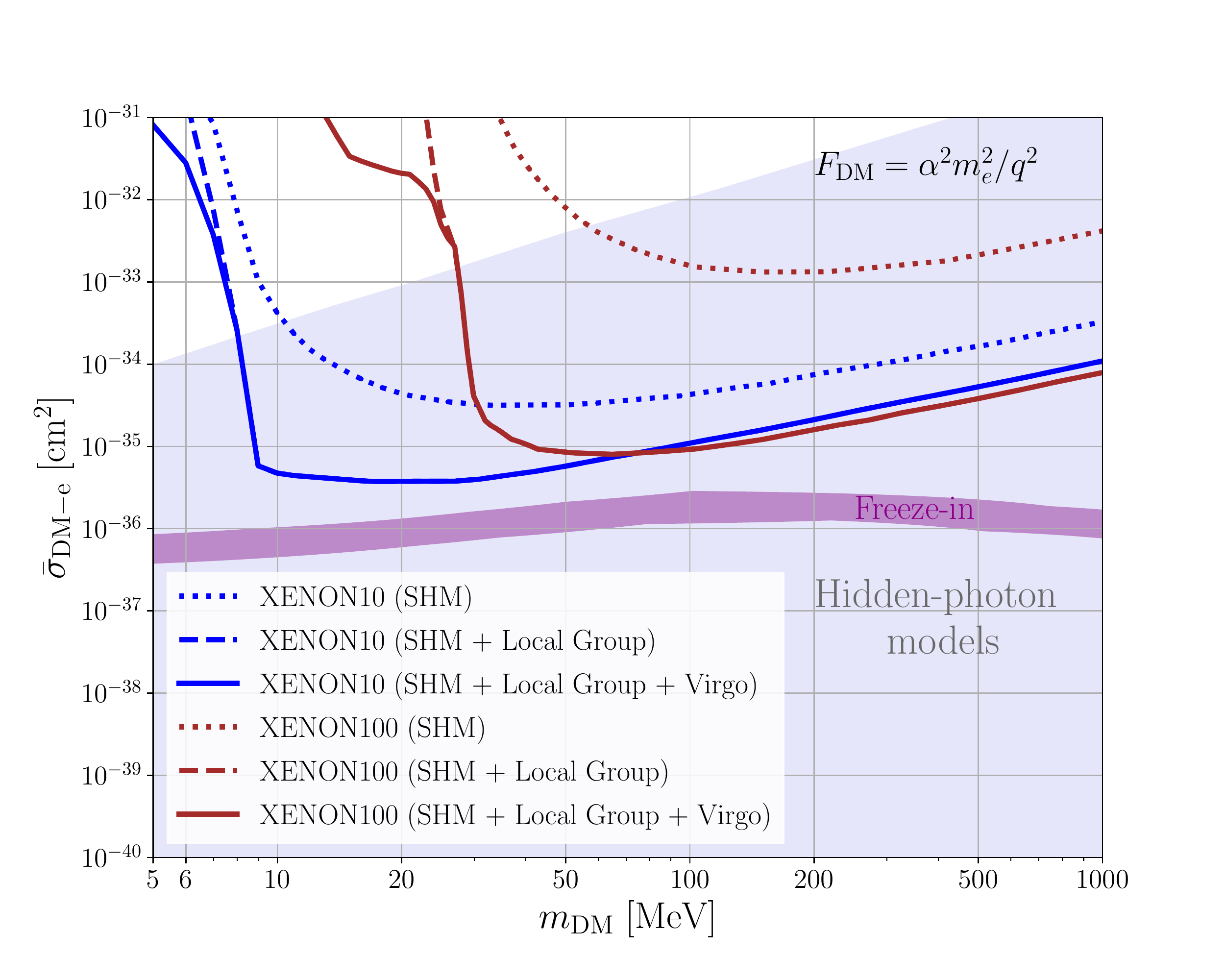}
\caption{\small Upper limits at the 90$\%$ C.L. on the dark matter-electron cross-section from the null search results from the XENON10 (blue) and XENON100 (brown) experiments, assuming  an interaction mediated by a heavy hidden photon (left panel) or an ultralight hidden photon (right panel) . 
The dotted line indicates the upper limit derived under the assumption that only galactic dark matter, described by the Standard Halo Model, contributes to the dark matter flux at the Solar System. The dashed lines show the impact of including in the flux also the  non-galactic dark matter component from the Local Group and the solid lines show the impact of including also the diffuse component of the Virgo Supercluster.
We also show in the shaded lavender region the values of parameters expected from some selected models (see main text for details).
}
\label{fig:e}
\end{figure}

Finally, we show in Figure \ref{fig:e} the $90\%$ C.L upper limits on the dark matter-electron scattering cross section at fixed momentum transfer $q=\alpha m_{e}$ from XENON10 and XENON100 data, including both the galactic and the non-galactic dark matter components, for a heavy mediator (left panel) and for an ultralight mediator (right panel). For $m_{\rm DM}=50-1000$ MeV, dark matter from the Local Group envelope significantly enhances the reach of the XENON100 experiment, by at least one order of magnitude, compared to the expectations of the Standard Halo Model. For $m_{\rm DM}=30-50$ MeV, close to the kinematical threshold of the XENON100 experiment, the enhancement is even more significant. Further, the non-galactic dark matter components allow to probe the mass region $m_{\rm DM}=13-30$ MeV, for which dark matter particles from the host halo do not induce detectable recoils. For the XENON10 experiment the conclusions are analogous, although in this case the enhancement is somewhat more modest, but still  ${\cal O}(1)$.  We also show in the plot values of parameters expected from theoretical models \cite{Essig:2012yx} for a heavy or an ultralight mediator, respectively. For the former, reproducing the correct thermal abundance via freeze-out requires values of $\bar{\sigma}_{\rm DM-e}$ above the purple line. For the latter, the purple shaded region shows the values favored by the freeze-in mechanism \cite{Essig:2017kqs}. Clearly, the non-galactic components significantly improve the discovery potential of experiments.

\section{Conclusions}
\label{sec:conclusions}

In this note, we have analyzed the impact of non-galactic dark matter in direct detection experiments, both for nuclear and for electron recoils. We have considered two possible non-galactic contributions to the dark matter population at the Solar System, stemming from the diffuse components of the Local Group and of the Virgo Supercluster. Their contribution to the local density is subdominant compared the one from the Milky Way halo. Yet, their contribution to the dark matter flux can be considerable at large velocities, and they can have a non-negligible impact in direct detection experiments. 

We have calculated the contribution from these two non-galactic contributions to the differential recoil rate.
Concretely, we have simulated the expected energy spectrum of nuclear recoils at the CRESST-III and the XENON1T experiment through the spin-independent interaction, and at CRESST-III, PICO-60 and XENON1T through the spin-dependent interaction. We have also simulated the expected energy spectrum of electron recoils at the XENON10 and XENON100 experiments through an ultralight or a heavy hidden photon. 

For nuclear recoils, we have found an enhancement in the energy recoil spectrum compared to the expectations from the Standard Halo Model, most notably close to the kinematical thresholds. Correspondingly, the upper limits on the scattering cross-section become more stringent for light dark matter. More concretely, for the spin-independent interaction, the limits for $m_{\rm DM}=1$ GeV (0.2 GeV) from CRESST-III  become a factor $\sim 2$ ($\sim 10^{3}$) more stringent than in the Standard Halo Model, and the limits for $m_{\rm DM}=10$ GeV (4 GeV) from XENON1T, a factor  $\sim 10$ ($\sim 10^{4}$). Similar conclusions hold for the spin-dependent interaction. Further, the dark matter mass range that experiments are able to probe is extended to lower values. We have also argued that the non-galactic dark matter component would leave a characteristic signature in the recoil spectrum in the form of step-like features, which could be discerned from the smooth spectrum expected from recoils induced by dark matter particles from the host halo or from the irreducible neutrino background. We also expect a non-negligible impact of the non-galactic contributions to the flux for inelastic dark matter scatterings, or  for secondary ionization induced by the Migdal effect. We leave this analysis for future work.  

For electron recoils, we find also an enhancement of the differential rate. Furthermore, the enhancement is appreciable over a larger range of recoil energies, and not only close to the kinematical thresholds. In turn, the limits on the dark matter-electron scattering cross-section are significantly strengthened in a wide mass range. For interactions mediated by a heavy hidden photon, the enhancement amounts to a factor of $\sim 2$ ($\sim 10$) at $m_{\rm DM}=1000$ MeV for the XENON10 (XENON100) experiment, and increases for lighter dark matter, being the enhancement a factor of $\sim 10^{2}$ at $m_{\rm DM}=10$ MeV for the XENON10 experiment and a factor of $\sim 10^{2}$ as well at $m_{\rm DM}=40$ MeV for the XENON100 experiment. For an ultralight mediator, the conclusions are analogous, being the strengthening of the limits somewhat larger for the XENON100 experiment. 

An obvious caveat of this analysis is the as yet poor understanding of the non-galactic dark matter phase-space distribution. Therefore, the limits on the cross-section derived in this work should be taken with a grain of salt. We hope that future astronomical observations, and a more refined modeling of the dark matter envelope of the Local Group and the Virgo Supercluster, will lead to a more robust assessment of the impact of these two contributions in direct dark matter searches. A proper understanding of the non-galactic components to the dark matter flux may prove to be crucial for the correct interpretation of the experimental data.

\section*{Acknowledgements}
This work has been supported by the Collaborative Research Center SFB1258 and by the Deutsche Forschungsgemeinschaft (DFG, German Research Foundation) under Germany's Excellence Strategy - EXC-2094 - 390783311. We are grateful to Dominik Fuchs, Federica Petricca and Tomer Volansky for very useful discussions and suggestions.

\appendix

\section{Derivation of limits from direct detection experiments}
\label{sec:details}
In this appendix, we include details with our procedure to calculate upper limits on the scattering cross-section from the experimental data. 

To derive upper limits on the SI and SD dark matter-nucleon cross section for CRESST-III \cite{Mancuso:2020gnm}, XENON1T \cite{Aprile:2018dbl} and PICO-60 \cite{Amole:2017dex}, we follow a poissonian-likelihood approach and use the detector response functions given in the  \texttt{DDCalc} package \cite{Workgroup:2017lvb,Athron:2018hpc}. 
For CRESST-III, we use the published data \cite{Abdelhameed:2019mac} corresponding to an exposure of 5.594 kg$\times$day, and we account for a finite energy resolution and cut-survival probability in the expected dark matter spectrum as described by the collaboration. All events in the acceptance region are considered signal events, which gives us a conservative $90\%$ C.L upper limit of 873.9 events. For the XENON1T experiment, we use the data from \cite{Aprile:2018dbl} with an exposure of 35.6  tonnes$\times$day. \texttt{DDCalc} divides the signal region into two energy bins, which correspond to [3,35] PE and [35,70] PE. The estimated background in both bins are 0.46 and 0.34 events, while the number of observed events are 0 and 2, respectively. The efficiencies were calculated simulating fluctuations of the S1 and S2 signal and using both scintillation and ionization yields. We consider a $90 \%$ C.L upper limit on the number of signal events of 3.9. Lastly, for the PICO-60 experiment, we use the results from \cite{Amole:2017dex}, corresponding to an exposure of 1167 kg$\times$day. Since PICO-60 observed no signal events, we take a $90\%$ C.L. upper limit on the number of signal events of 2.3.

For the calculation of the ionization rate we follow \cite{Essig:2012yx} to model the conversion from the electron's recoiling energy $E_{er}$ to the experimental observable at the XENON1T experiment, the number of photoelectrons (PE). The recoiling electron will ionize and excite other atoms, yielding Floor$(E_{er}/W)$ primary quanta in form of observable electrons $n_{e}$ and unobservable photons. We take the value of the average energy needed to produce a single quanta (photon or electron) to be $W$=13.8 eV. Further, we choose the probability for the initial electron to recombine with an ion to be zero and the fraction of primary quanta observed as electrons to be $0.83$.
 A more refined modeling of the electron ionization and the associated uncertainties at XENON1T can be found in \cite{Aprile:2019xxb,Essig:2017kqs, Baxter:2019pnz}.

We then calculate $90\%$ C.L upper limits on the dark matter-electron scattering cross section at fixed momentum transfer $q=\alpha m_{e}$ using XENON10 and XENON100 data.
The experiments report the number of photoelectrons (PE) produced by an event. To convert the $n_{e}$ into PE, we assume that an event with $n_{e}$ electrons produces a gaussian distributed number of PE with mean $n_{e}\mu$ and width $n_{e}\sigma$, where $\mu=27(19.7)$ and $\sigma=6.7(6.2)$ for XENON10 (XENON100). We consider the energy range in XENON10 going from 14 to 95 PE, corresponding up to $n_{e}=3$. For XENON100, we consider the energy range going from 80 to 110 PE, corresponding to $n_{e}$=4 and $n_{e}$=5. We use the binned $90\%$ C.L. upper bounds on the event rate calculated in \cite{Essig:2017kqs}, obtained after multiplying the signal with the trigger and acceptance efficiencies. We notice that our limits for the SHM flux are more conservative than those of \cite{Essig:2017kqs} . This is likely due to the fact that we are considering only the three outermost orbitals of xenon (5p, 5s and 4d), while the referenced work considers the orbitals 4s and 4p as well. Furthermore, the energy thresholds considered in this note for XENON10 and XENON100 are $n_{e}=3$ and $n_{e}=5$, while \cite{Essig:2017kqs} considers $n_{e}$ up to 6 in both cases. We furthermore point out that the XENON1T collaboration has provided upper limits on the dark-matter electron scattering cross section at fixed momentum transfer for the heavy mediator scenario $F_{\rm DM}=1$ \cite{Aprile:2019xxb}, and these are also stronger than ours at $m_{\rm DM} \gtrsim$ 30 MeV, as well as those from \cite{Essig:2017kqs}. Again, the energy range considered in this search and different exposure explains the gap with our results. \\

\bibliographystyle{JHEP-mod}
\bibliography{references}

\end{document}